\documentclass[aps,prl,preprint,groupedaddress]{revtex4-1}

\usepackage{graphicx}
\usepackage{color}
\usepackage{dcolumn}
\usepackage{amsmath}
\usepackage{amsfonts}
\usepackage{bm}
\usepackage{epstopdf}
\usepackage{picture}

\usepackage{dcolumn}
\usepackage{amsmath}
\usepackage{amssymb}
\usepackage{amsfonts}
\usepackage{bm}
\usepackage{picture}

\DeclareMathOperator{\erfc}{erfc} 

\newcommand{\trq}{\Gamma}

\begin{document}

\title[Entropy production in micro-machines]{Entropy production in an elementary, light driven micro-machine} 

\author{Stuart J. Box\,$^{1}$, Michael P. Allen\,$^{1,2}$, David B. Phillips\,$^{3}$ and Stephen H. Simpson\,$^{4,*}$}
\affiliation{%
$^{1}$H.H. Wills Physics Laboratory,  University of Bristol, Tyndall Avenue, Bristol BS8 1TL, UK \\
$^{2}$Department of Physics, University of Warwick, Coventry CV4 7AL, UK \\
$^{3}$Department of Physics and Astronomy, University of Exeter, Exeter, EX4 4QL. UK \\
$^{4}$Institute of Scientific Instruments of the Czech Academy of Science, v.v.i.,  
Kr\'alovopolsk\'a 147, 612 64 Brno, Czech Republic.}

\begin{abstract}
We consider the basic, thermodynamic properties of an elementary micro-machine operating at colloidal length scales. In particular, we track and analyse the driven stochastic motion of a carefully designed micro-propeller rotating unevenly in an optical tweezers, in water. In this intermediate regime, the second law of macroscopic thermodynamics is satisfied only as an ensemble average, and individual trajectories can be temporarily associated with decreases in entropy. We show that our light driven micro-propeller satisfies an appropriate fluctuation theorem that constrains the probability with which these apparent violations of the second law occur. Implications for the development of more complex micro-machines are discussed. 
\end{abstract}

\maketitle

\section{Introduction}
Advances in micro-fabrication techniques enable the manufacture of finely structured objects with ever greater precision \cite{LaFratta2017Twophoton}. As a consequence, increasingly sophisticated micro-machines are being developed \cite{Knopf2018Light,Xu2018Micro,Andrew2020Optical} to perform a variety of previously unrealisable tasks in the colloidal regime. Examples include the controlled transport, micro-fluidics, pumping and mixing \cite{Hong2010Light,Chen2018Light,Huang20153D,Swartzlander2011Stable,Simpson2012Optical}, sensing \cite{Phillips2014Shape}, or even hydrodynamic manipulation \cite{Butaite2019Indirect}. More recently, self-organising,  dynamically reconfigurable and self-propelled micro-machines have been exhibited \cite{Cui2019Nano,Alapan2019Shape,Huang2016Soft,Tottori2012Magnetic}.  
Quantitatively describing the behaviour of machines at these length scales is not trivial. Classical thermodynamics was developed, in part, to help optimize the efficiency of conventional machines. This framework, however, is inadequate for the colloidal regime where the energy flows that drive the machines are comparable in size to stochastic thermal fluctuations. In this article, we analyse the influence of thermal fluctuations on an elementary micro-machine, in this case a light driven micro-propeller or light-mill, observing that they are constrained by an appropriate \textit{fluctuation theorem} \cite{Evans2002Fluctuation}. In doing so, we underscore the usefulness of applying the principles of stochastic thermodynamics \cite{Seifert2012Stochastic} to the design of novel micro-machines.

\noindent As with biological analogues such as molecular motors \cite{Hoffmann2016Molecular}, artificial micro-machines operate in an intermediate thermodynamic regime. Conventional machines, working at every-day length scales, conform to the laws of classical, macroscopic thermodynamics where physical processes are strictly irreversible and associated with increasing disorder, or entropy. In contrast, dynamical motion at microscopic length scales is governed by time symmetric equations of motion. The thermodynamics of micro-machines are neither purely reversible nor completely irreversible and the second law of thermodynamics, which prohibits entropy from decreasing with time, appears not as an absolute law, but as a limiting case, emerging only for sufficiently large systems, over sufficient time intervals \cite{Seifert2012Stochastic,Evans2009Dissipation}. For micro-machines, working in the colloidal regime, individual trajectories can be associated with decreasing entropy for finite periods of time. The relative probability of observing such a trajectory is constrained by the fluctuation theorem (FT), the term really applying to a family of theorems, and decreases exponentially with time, with the second law restored in the limit either of long times or large system sizes. For a general system the FT takes the following form:\\
\begin{equation}\label{eq:dft0}
\frac{P(\Sigma_t=+\Sigma)}{P(\Sigma_t=-\Sigma)}=\exp(\Sigma),
\end{equation}
where $\Sigma_t$ is the entropy production of a trajectory integrated over time $t$ 
(divided by Boltzmann's constant $k_{\mathrm{B}}$ to make it dimensionless) 
and $P(\Sigma_t=\Sigma)$ the probability density that $\Sigma_t$ takes the value $\Sigma$. Equation (\ref{eq:dft0}) therefore quantifies the relative probabilities of entropy producing and consuming trajectories. Since $\Sigma_t$ is extensive, it increases with the system size and averaging time. An immediate consequence of Eq.~(\ref{eq:dft0}) is the \textit{second law inequality}, $\langle \Sigma_t \rangle \ge 0$, where the average is taken over an ensemble of experiments with the same start time, averaged over a fixed time period $t$ \cite{Evans2009Dissipation}. The second law inequality requires that the average entropy change is non-decreasing. Unlike the classical second law, it does not, however, prohibit decreases in entropy occurring in some of the realizations comprising the ensemble. 

\noindent Equation (\ref{eq:dft0}) is commonly referred to as the transient or detailed FT (DFT) to distinguish it from the integrated version, 
\begin{equation}\label{eq:ift0}
\frac{P(\Sigma_t<0)}{P(\Sigma_t>0)}=\Big \langle \exp(-\Sigma_t) \Big \rangle_{\Sigma_t>0},
\end{equation}
where $P(\Sigma_t<0)=\int^0_{-\infty} d\Sigma_t P(\Sigma_t)$ etc.\ and $\langle\ldots \rangle_{\Sigma_t>0}$ denotes an average over all trajectories for which $\Sigma_t>0$. Again, since $\Sigma_t$ is extensive, the right hand side of Eq. (\ref{eq:ift0}) approaches zero as the system size or time duration increase.\\\\
We wish to confirm appropriate versions of the fluctuation theorems, Eqs.~(\ref{eq:dft0},\ref{eq:ift0}) for an elementary micro-machine. Previous tests of the FT involve dragging spherical beads through water with optical traps \cite{Reid2004Reverse,Carberry2004Fluctuations} under various conditions \cite{Speck2007Distribution}. The study presented below extends these experiments to a more complex geometry. As will become clear, this increase in complexity gives rise to a number of experimental challenges, especially those relating to the design of our micro-machine and to the tracking of its motion.

\noindent Our chosen machine is a simple, light driven propeller or light-mill, which we fabricate using two-photon polymerization \cite{Cumpston1999Twophoton}. We track its motion as it rotates in a laser trap, showing that it conforms to an appropriate FT. The propeller has been carefully designed so that it rotates slowly about its axis, allowing its motion to be tracked with sufficient accuracy and resolution to observe trajectories with temporarily decreasing entropy. Small, natural asymmetries in the system cause the axial torque to vary strongly with orientation, so that the rotational motion is highly uneven. Rather than being a defect of the study, these irregularities are an important part of it. The case of constant torque is comparatively trivial. In particular, diffusion of a particle exposed to a spatially invariant force or torque is very well understood \cite{Coffey1996Langevin}. In our case, the dependence of the torque on the orientation of the propeller provides a more demanding and interesting test of the FT.

\noindent In the following sections we review the necessary theoretical background, which closely follows the work of Speck, Seifert et al. \cite{Speck2007Distribution}, describe the design and fabrication of the rotor and the experimental procedure. We next present tests of the detailed and integrated forms of the rotational FT.  We comment on the small discrepancies between theory and experiment  and conclude with a discussion of the implications for the optimal design of future micro-machines.  

\section{Theoretical background}
We consider the over-damped motion of a Brownian propeller or rotor, rotating about a fixed axis in water. This motion is described by the Langevin equation,
\begin{equation}\label{eq:Lang0}
\xi_{\mathrm{r}} \dot \phi = \trq(\phi)+\trq^{\mathrm{S}}(t),
\end{equation}
where $\phi$ is the orientation of the rotor and $\xi_{\mathrm{r}}$ is the rotational friction. The systematic, external torque, $\trq(\phi)$ varies with $\phi$ but is always of the same sign and $\trq^{\mathrm{S}}(t)$ is the stochastic, Langevin torque with zero mean, uncorrelated and with variance determined by the fluctuation dissipation theorem i.e.,
\begin{eqnarray}\label{eq:Tfluct}
\langle \trq^{\mathrm{S}}(t) \rangle &=& 0 \\
\langle \trq^{\mathrm{S}}(t) \trq^{\mathrm{S}}(t') \rangle &=& 2k_{\mathrm{B}}T \xi_{\mathrm{r}} \delta(t-t').
\end{eqnarray}
Thus, we focus on axial rotations, assuming that the motion of all other degrees of freedom are negligible or independent. Qualitative features of the motion follow from a consideration of the impulses, $I$, delivered to the rotor over a finite time interval, $\Delta t$ i.e.\  $I=\int^{\Delta t} dt \trq(t)$. For short time intervals, the impulse from the systematic torque, $I^{\mathrm{sys}}$, is obviously proportional to $\Delta t$ i.e.\ $I^{\mathrm{sys}} \approx \trq(\phi_0) \Delta t$, if $\phi_0$ is the initial orientation and we assume the torque does not vary too much over $\Delta t$. By comparison, the impulse from the stochastic torque, $I^{\mathrm{stoch}}$, is a Gaussian random variable drawn from a distribution with a variance of  $\langle (I^{\mathrm{stoch}})^2 \rangle = 2k_{\mathrm{B}}T \xi_{\mathrm{r}} \Delta t$. The probability that the stochastic impulse exceeds the systematic impulse is therefore, 
\begin{equation}\label{eq:imp0}
P(|I^{\mathrm{stoch}}| > |I^{\mathrm{sys}}|) \approx \erfc \Bigg(\frac{\trq(\phi_0)\sqrt{\Delta t}}{2\sqrt{k_{\mathrm{B}}T\xi_{\mathrm{r}}}} \Bigg).
\end{equation}
In other words, for very small $\Delta t \rightarrow 0$ we expect the total impulse to be dominated by the stochastic term i.e.\ $I\equiv I^{\mathrm{stoch}}+I^{\mathrm{sys}} \approx I^{\mathrm{stoch}}$. It may therefore act in the opposite sense to the systematic torque. As $\Delta t$ increases, this becomes progressively less likely, until the total impulse is approximately equal to the contribution from the systematic torque. The angular displacements made by the rotor may be expected to behave similarly i.e.\ for short times there is significant probability that the propeller rotates in the opposite sense to the applied torque while, for greater time intervals, the propeller is practically guaranteed to rotate with the systematic torque. Furthermore, if we consider the probability, $p(\phi,t)d\phi$ that, at time $t$, the propeller has an orientation in the interval $[\phi, \phi+d\phi]$, it is clear that $p(\phi,t)$ must approach a steady state limit, $p_{\mathrm{ss}}(\phi)$, as the time interval over which data is collected increases i.e.\ $\lim_{t\rightarrow \infty} p(\phi,t) = p_{\mathrm{ss}}(\phi)$. The time evolution of $p(\phi,t)$ is determined by the Fokker-Plank equation with drift and diffusion coefficients derived from the Langevin equation, Eq. (\ref{eq:Lang0}), 
\begin{equation}\label{eq:fp0}
\partial_tp(\phi,t)=-\partial_{\phi}j(\phi,t)=-\frac{1}{\xi_{\mathrm{r}}}\partial_{\phi} \Big[ \trq(\phi)p(\phi,t)-k_{\mathrm{B}}T\partial_{\phi}p(\phi,t) \Big],
\end{equation}
where $p(\phi,t)$ is the probability distribution of the orientation and $\xi_{\mathrm{r}} j(\phi,t)=\trq(\phi)p(\phi,t)-k_{\mathrm{B}}T\partial_{\phi}p(\phi,t)$ is the total probability current, the first and second terms of which correspond, respectively, to the drift and diffusion currents. As time progresses, $p(\phi,t)$ approaches a non-equilibrium steady state, $p_{\mathrm{ss}}(\phi)$. Since we are treating this as a one dimensional system, the associated steady state current, $j_{\mathrm{ss}}$ and mean angular velocity, $\omega_{\mathrm{ss}}(\phi)$ satisfy $j_{\mathrm{ss}}=p_{\mathrm{ss}}(\phi)\omega_{\mathrm{ss}}(\phi)$, where $j_{\mathrm{ss}}$ is independent of orientation in accordance with conservation of probability i.e.
\begin{equation}\label{eq:jss}
j_{\mathrm{ss}}=\frac{1}{\xi_{\mathrm{r}}}\Big[\trq(\phi)p_{\mathrm{ss}}(\phi)-k_{\mathrm{B}}T \partial_{\phi}p_{\mathrm{ss}}(\phi) \Big]=p_{\mathrm{ss}}(\phi)\omega_{\mathrm{ss}}(\phi).
\end{equation}
This system has several unique features. Since it is clearly not at equilibrium, and the torque is not purely conservative, we might expect thermodynamic properties, such as the work performed by the propeller, to depend on the detailed path taken, $\phi(t)$ for $t_0\leq t \leq t_1$, not purely on the end points, $\phi(t_0)$ and $\phi(t_1)$. However, this is not quite correct. $\trq(\phi)$ is a function of a single coordinate only. As such, it can be written as the gradient of a scalar function, providing we extend the range of $\phi$ to cover the real line. We therefore have a non-conservative system in which, unusually, we should be able to evaluate path dependent integrals in terms of end point values. For example, if we write $\trq(\phi)=T^0+T^1(\phi)$, where $T^0$ is the average value of $\trq(\phi)$ in the range $[0,2\pi)$, then the corresponding potential is, 
\begin{eqnarray}\label{eq:V0}
V(\phi^{\mathrm{c}})=-\int^{\phi^{\mathrm{c}}}_0 d\phi' \trq(\phi')&=&-T^0\phi^{\mathrm{c}}-\int^{\phi^{\mathrm{c}}}_0d\phi'T^1(\phi')\nonumber \\
&=&-T^0\phi^{\mathrm{c}}-\int^{\phi}_0d\phi'T^1(\phi') \nonumber \\
&\equiv&-T^0\phi^{\mathrm{c}}+V^0(\phi).
\end{eqnarray}
Here, $\phi^{\mathrm{c}}$ is a continuously varying angular displacement that can become arbitrarily large and counts multiple complete rotations i.e. it is the total angular displacement. The upper limit in the integral appearing in the final term of Eq. (\ref{eq:V0}) can be changed from $\phi^{\mathrm{c}}$, which measures the total angular displacement, to $\phi$, the orientation of the propeller (i.e.\ $\phi = \phi^{\mathrm{c}}-2\pi n$ for integer  $n=\lfloor\phi^{\mathrm{c}}/2\pi\rfloor$) since the integral of $T^1(\phi')$ over a complete rotation is zero by construction. Equation~(\ref{eq:V0}) reveals the anomalousness of the system. If we take the system around an arbitrary closed path, $\phi(t)$, so that it returns to its initial orientation, the total work done does not depend on the details of the trajectory, only on the integer number of complete cycles traversed, i.e.\ $w=\int dt \trq(\phi)\dot\phi=2\pi n T^0$. If the system is constrained somehow, so that no complete revolutions are executed, it is effectively conservative. Nevertheless, the fact that the torque can be expressed as the derivative of a scalar function implies that relevant path dependent integrals can always be evaluated in terms of their end points, ignoring the details of the stochastic motion. To illustrate, the dissipated heat is \cite{Speck2007Distribution},
\begin{equation}\label{eq:heat0}
dq=\trq(\phi)d\phi \qquad q[\phi(\tau)]=\int^t_0 d\tau \trq(\phi(\tau)) \dot \phi(\tau)=T^0 \phi^{\mathrm{c}} - \Delta V^0,
\end{equation}
where the integral in Eq.~(\ref{eq:heat0}) is evaluated by writing the torque as the derivative of the potential in Eq.~(\ref{eq:V0}), and $\Delta V^0=V^0(\phi(t))-V^0(\phi(0))$ and $\phi$ again, is the orientation whereas $\phi^{\mathrm{c}}$ is the total angular displacement.
Following \cite{Speck2007Distribution}, the total change in entropy along a trajectory is given by the sum of the change in the medium,
\begin{equation}\label{eq:smed}
\Delta \Sigma_{\mathrm{m}}[\phi(\tau)] = \frac{\Delta s_{\mathrm{m}}[\phi(\tau)]}{k_{\mathrm{B}}}=\frac{q[\phi(\tau)]}{k_{\mathrm{B}}T},
\end{equation}
and the change in the system entropy,
\begin{eqnarray}\label{eq:ssys}
\Delta \Sigma(x_0,x_t) = \Sigma(x_t)-\Sigma(x_0), \nonumber \\
\Sigma(x_\tau)=\frac{s(\tau)}{k_{\mathrm{B}}}=-\ln p_{\mathrm{ss}}(\phi),
\end{eqnarray}
where $p_{\mathrm{ss}}(\phi)$ is the steady state distribution of the propeller orientation. This latter quantity, the system entropy, is defined in analogy with the usual Gibbs entropy, with the integral over microstates replaced by an integral along the particle trajectory \cite{Seifert2005Entropy}. Since we have assumed that the sign of the torque is the same for all orientations (a condition satisfied for the propeller used in the experiments, see below), the total potential (Eq.~(\ref{eq:V0})) is monotonic. Trajectories associated with the consumption of medium entropy (Eq.~(\ref{eq:smed})) therefore relate to small rotations against the applied torque which, as discussed above, occur only over short time intervals. We note that Eq.~(\ref{eq:heat0}) can be arrived at by various other routes. For instance, the torque can be separated into a part derived from a $2\pi$ periodic potential, $V^0$ say, and a constant, non-conservative term. By computing the work from an integral of the non-conservative torque, and the change in internal energy from $V^0$, an application of the integrated first law reproduces Eq.~(\ref{eq:heat0}) \cite{Speck2007Distribution}. Of course, the force can be partitioned into conservative and non-conservative components in infinitely many ways, all leading to the same result. As stated above, the fundamental reason is that this is a one dimensional system so the force can be represented as the derivative of a scalar function, as in Eq.~(\ref{eq:V0}).

\noindent The principles described above provide us with a simple way of testing the rotational fluctuation theorem for a driven micro-propeller. First we measure the steady state orientation distribution, $p_{\mathrm{ss}}(\phi)$ and average rotational velocity, $\omega_{\mathrm{ss}}(\phi)$. Next we evaluate the systematic torque from the steady state probability current, $j_{\mathrm{ss}}$, Eq.~(\ref{eq:jss}). This step requires the rotational drag, $\xi_{\mathrm{r}}$, which we estimate numerically \cite{Carrasco1999Hydrodynamic}. We note that it is not possible to find an exact, independent value for $\xi_{\mathrm{r}}$ since the precise dimensions of the micro-fabricated rotor are not available. The numerical estimate of $\xi_{\mathrm{r}}$ is therefore treated as a guide and, since it is a scalar, it is straightforward to consider moderate variations in this parameter. Finally, Eqns.~(\ref{eq:smed}),(\ref{eq:ssys}) are used to evaluate entropy changes over time intervals of varying length and apply them in the integrated and detailed fluctuation theorems, Eq.~(\ref{eq:dft0}),(\ref{eq:ift0}). Achieving this experimentally is rather demanding, and the protocols and techniques we adopted are described in greater detail below.
\section{Propeller design and fabrication}
Testing the rotational FTs described above places some demanding constraints on the design and behaviour of our micro-propeller. It must exhibit three key characteristics. \textbf{(1)}, the propeller should rotate about a single fixed axis, with minimal fluctuations in the orientation of the axis. \textbf{(2)}, it should rotate slowly enough for the short term, entropy consuming trajectories to be accurately resolved by the available technology. Finally, \textbf{(3)} its shape should facilitate accurate tracking of its orientation. We discuss each of these issues in more detail below.

\begin{figure}[h!]
\begin{center}
\includegraphics[width=5cm]{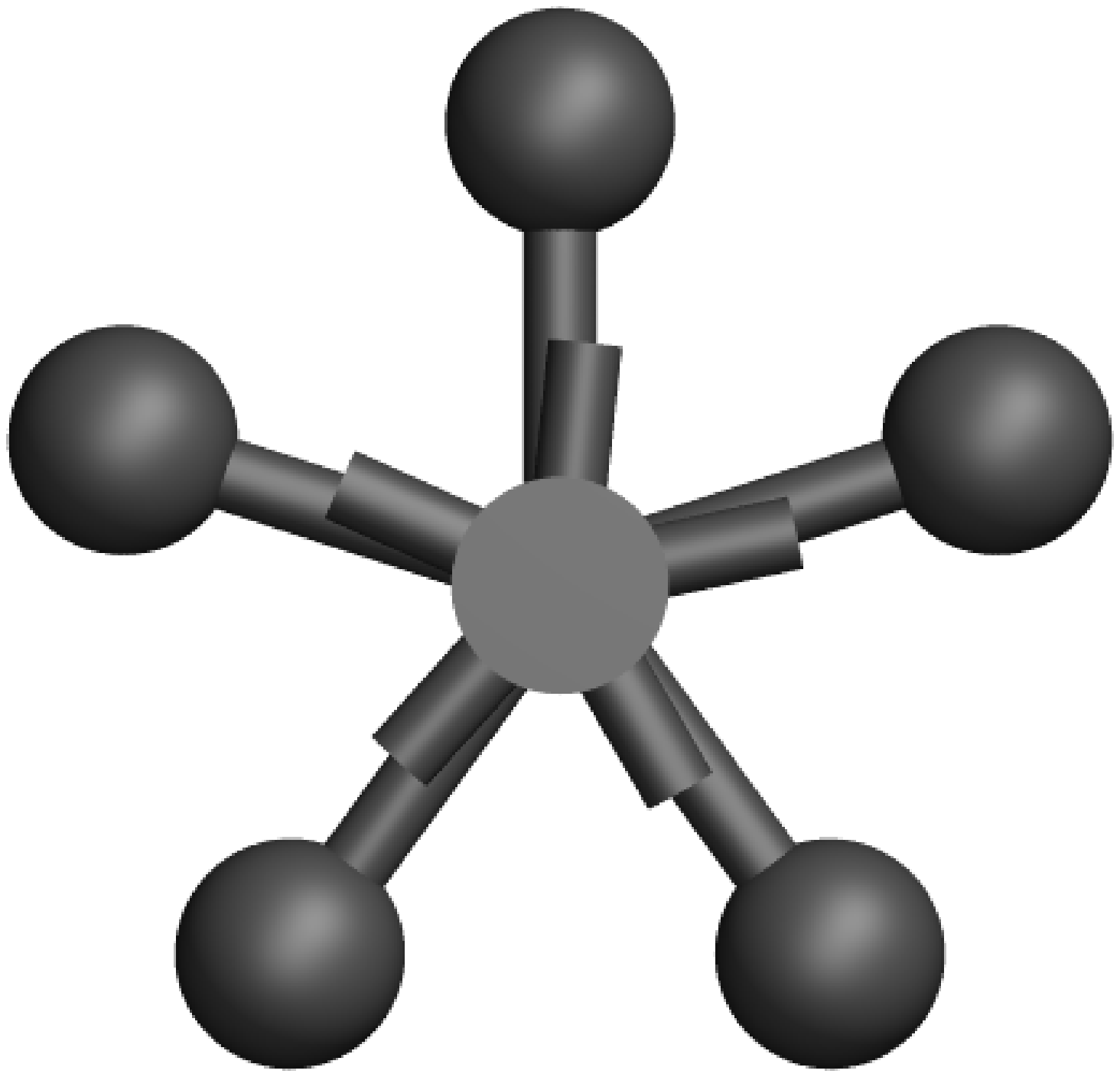}
\includegraphics[width=5cm]{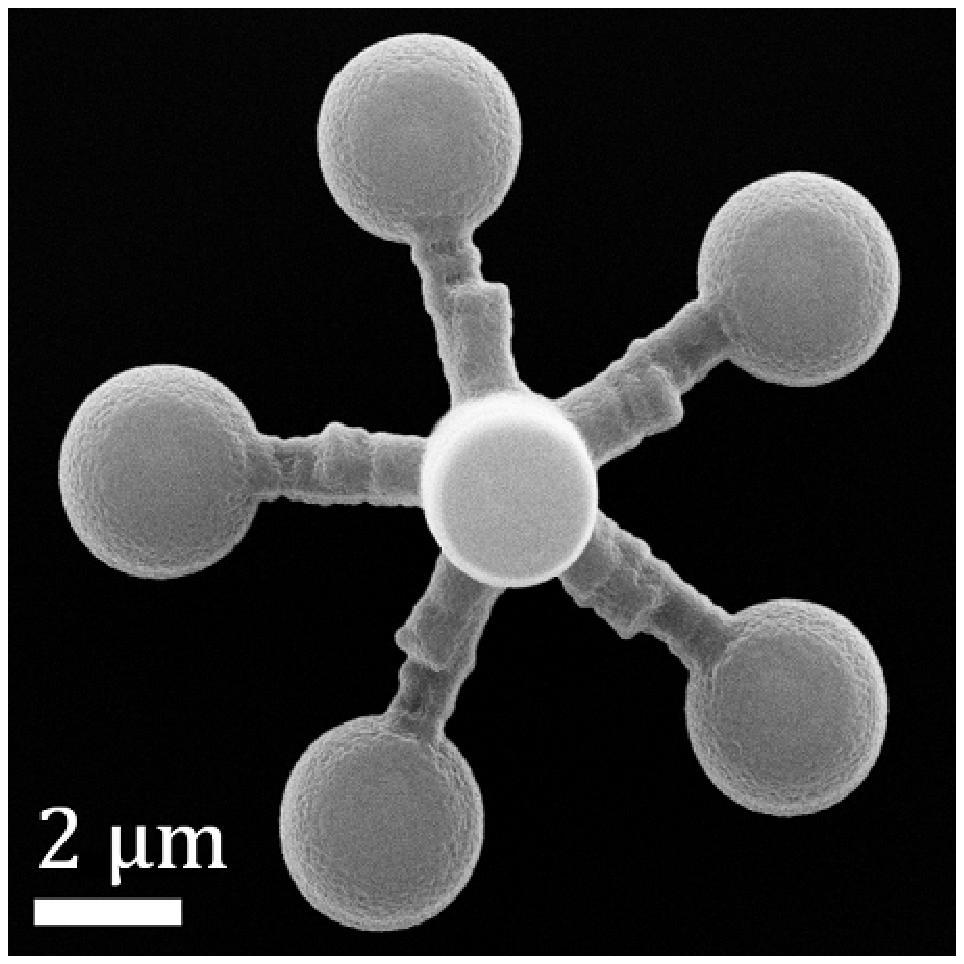}
\end{center}
\caption{The micropropeller used in the experiment. (Left) Schematic of the design, rendered using POV-Ray, (Right) SEM image of the microfabricated (two-photon polymerization) propeller used in the experiment.}\label{fig:prop}
\end{figure}

\noindent \textbf{(1) Axis stability: } Elongated, dielectric objects tend to align themselves with the axis of optical beams \cite{Simpson2012Stability}. The stability of the rotation axis of a micro-propeller is therefore promoted by providing it with a long, sturdy central spindle. \\
\textbf{(2) Rotation rate: } Equation (\ref{eq:imp0}) allows us to find a crude estimate of the rotation rate required to observe entropy consuming trajectories i.e.\ we require $\Delta t \lessapprox 4k_{\mathrm{B}}T\xi_{\mathrm{r}} / \trq^2(\phi_0) = 4k_{\mathrm{B}}T/\xi_{\mathrm{r}} \omega^2$, where $\omega$ is the typical angular velocity at $\phi_0$. Taking $\Delta t_{\mathrm{min}}$ to be the smallest time interval over which we can accurately measure gives $\omega^2 \approx 4k_{\mathrm{B}}T/\xi_{\mathrm{r}} \Delta t_{\mathrm{min}}$. Finally, if we approximate the rotational drag by that of a sphere, $\xi_{\mathrm{r}} \approx 8 \pi \mu a^3$ with an effective radius $a=2\mu$m we get $\omega \lessapprox 0.1 / \sqrt{\Delta t_{\mathrm{min}}}$. In our case, 
$\Delta t_{\mathrm{min}} = $~1~ms, 
suggesting that we require a rotation rate of less than 0.5~Hz. Most light-mills rotate at greater rates, typically in excess of several Hz \cite{Galajda2002Rotors,Asavei2009Optical,Brzobohaty2015Complex}, and more effort is devoted to increasing rotation frequency than decreasing it \cite{Loke2014Driving}. Designing a propeller that rotates this slowly without stalling completely is a greater experimental challenge than might have been expected. To meet it, we recall the basic symmetry properties of optical torque coupling \cite{Simpson2007Polarization}. An object with rotational symmetry and a mirror plane that includes the beam axis has a preferred orientation in a linearly polarized optical tweezer, while an object with chiral symmetry experiences a constant, orientation independent torque.
We adopt a basic design for the propeller in which the torque it experiences should be controllable through a single parameter.
This design consists of a propeller with two tiers of five cylindrical arms. When the tiers are aligned, the rotor has a mirror plane, and does not rotate. By gradually displacing the tiers with respect to one another, we can control the rotation speed. Natural imperfections in the beam and propeller result in a torque that, rather than being constant, is strongly dependent on orientation. This tendency becomes more conspicuous the closer the propeller is to having a mirror plane. \\
\textbf{(3) Tracking: } Finally, it must be possible to track the orientation of the propeller with as much angular precision as possible. To do this, we add microspheres to the ends of the rotor arms, and make use of the established methods for high precision sphere tracking \cite{Bowman2010Particle}. The orientation of the propeller can then be accurately retrieved from measurements of the sphere positions. With these principles in mind, we arrived at the following final design, shown in Fig. (\ref{fig:prop}). 

\noindent We fabricate micro-propellers using two-photon polymerization. To implement the geometric design shown in Fig.~(\ref{fig:prop} left), we use custom software (LabVIEW) to generate an instruction file to control the path of a laser beam in a two-photon polymerization machine (Photonic Professional, Nanoscribe). By specifying the beam path directly, rather than relying on slicing software to automatically generate a path from a CAD file, we retain fine control over the fabrication process. We deposit approximately 20~$\mu$L of photoresist material (IP-G 780, Nanoscribe) on a coverslip, and then heat it using a hotplate (100{}C for 1 hour), following the manufacturer's instructions for photoresist preparation. We then fabricate many ($>$50) copies of our micro-propeller design using the two-photon polymerization machine to scan a laser beam through the prepared photoresist according to our custom beam path file. The polymerised material is developed, and unpolymerised material washed away, by submersing the coverslip first in developing fluid (Microposit EC Solvent) for 20 minutes, followed by isopropanol for 5 minutes. We move the fabricated micro-propellers into suspension by placing a drop of water (Direct-Q 3 UV, Millipore) on top of them, and then apply mechanical agitation to detach them from the coverslip using a thin wire manipulated by a 3-axis translation stage (ULTRAlign Precision XYZ Linear Stage, Newport), similar to a previously employed method~\cite{Phillips2012Force,Phillips2013fashioning} . The suspension is transferred to a custom microscope sample chamber consisting of one microscope slide and three coverslips bonded together (UV Adhesive 81 or 68, Norland) using a pipette, and the remaining volume of the chamber is filled with additional water then sealed with adhesive. This process typically results in around 50-80\% of the fabricated micro-propellers being transferred into the microscope sample chamber. The two-photon polymerization and transfer steps are all carried out in a clean room.

\section{Experiment}

We optically trap one of the micro-propellers inside the sample chamber using a custom-built holographic optical tweezers system, which is similar to a system described elsewhere~\cite{Gibson2008Holographic}. The optical tweezers are based around a 1070~nm laser (YLM-5-LP-SC, IPG Photonics) whose beam is expanded to fill a spatial light modulator, or SLM (XY Series HSP512-1064-DVI, Boulder Nonlinear Systems). The beam is then tightly focussed using an oil immersion, 1.4~NA, 100x objective lens (Plan-Apochromat, Zeiss) to form optical trap(s). Relative translation of the sample and optics is achieved with a piezo-based translator (Mipos 140 PL, Piezosystem Jena) along the optical ($z$) axis, and with a servomotor-based stage (MS2000, ASI) in the orthogonal ($xy$) plane. Imaging of the sample is performed by illumination from a custom LED source (University of Glasgow). Light from the LED which is transmitted by the sample is then collected by the objective lens and directed to a camera (EoSens CL MC1362, Mikrotron) using a polarising beam-splitter. Images from the camera are acquired on a computer through a PCI-e frame grabber (PCIe-1433, National Instruments). We use the ``Red Tweezers" software~\cite{Bowman2010Particle} to generate and position optical traps through control of the SLM. A single micro-propeller is picked up with an optical trap, such that the long axis of the propeller's spindle aligns with the optical axis. The micro-propeller is then translated to the center of the sample chamber in the $xy$ plane, and to a distance of around 30~{$\mu$}m above the coverslip along the $z$ axis using the microscope stages. Note that the latter distance is limited by the working distance of the objective lens. The micro-propeller is simultaneously confined in translational space and driven to rotate continuously about its spindle by the optical trap.

We record a high speed (1~kHz) video of a trapped and rotating micro-propeller using a custom LabVIEW program to acquire images from the system's camera. The fast frame rate is achieved by; (i) reducing the active area of the camera to the minimum needed to view the propeller, (ii) use of a signal generator (33220A, Agilent) to trigger the frame grabber, and (iii) use of a circular buffer in software that losslessly transfers acquired images to an SSD drive (HyperX SH100S3/120G, Kingston). In our set up, the frame rate is limited by illumination intensity and control over the camera's exposure time.

\noindent Data analysis is performed offline, starting by extracting the position of the micro-propeller in each frame of the video using an additional custom LabVIEW program. We first find the Cartesian position of each of the five spheres on the micro-propeller, using a symmetry transform tracking library, taken from the Red Tweezers software~\cite{Bowman2010Particle,Bowman2014Red}. The propeller's orientation, $\phi$, can then be defined by the Cartesian position of one of the spheres and the position of the propeller's center. The center is calculated in each frame as the average of the positions of all five spheres.

\begin{figure}[h!]
\begin{center}
\includegraphics[width=12cm]{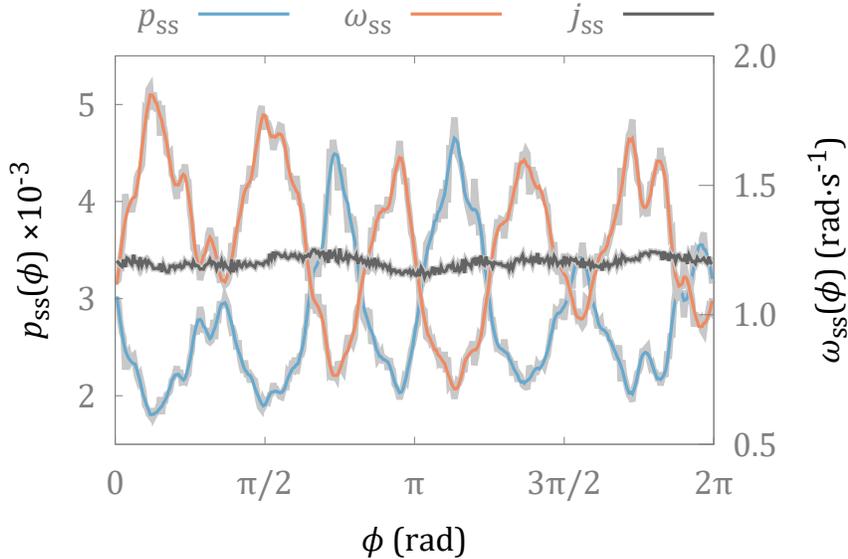}
\end{center}
\caption{Steady state distributions of the orientation, $p_{\mathrm{ss}}(\phi)$, angular velocity, $\omega_{\mathrm{ss}}(\phi)$ and probability current, $j_{\mathrm{ss}}$. In each case, the raw data (grey) is overlayed with the smoothed data. The scale used for $j_{\mathrm{ss}}$ is the numerically equivalent to that used for $p_{\mathrm{ss}}(\phi)$, with units rad/s.}\label{fig:steady}
\end{figure}

\begin{figure}[h!]
\centering
\begin{tabular}{cc}
\includegraphics[width=8cm]{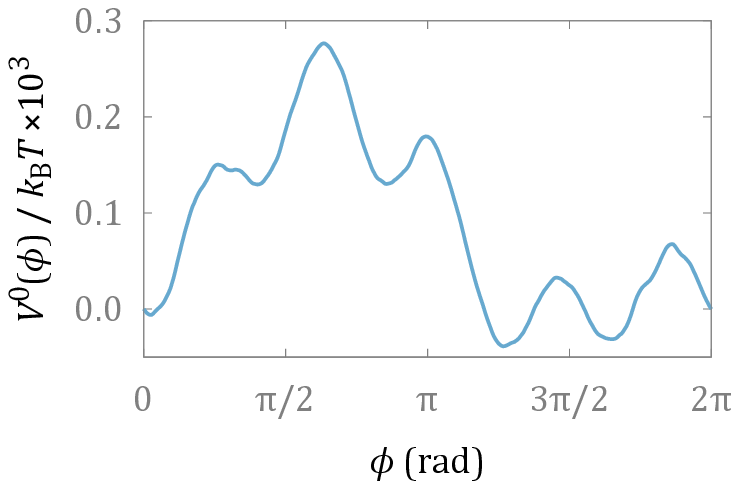}
\includegraphics[width=8cm]{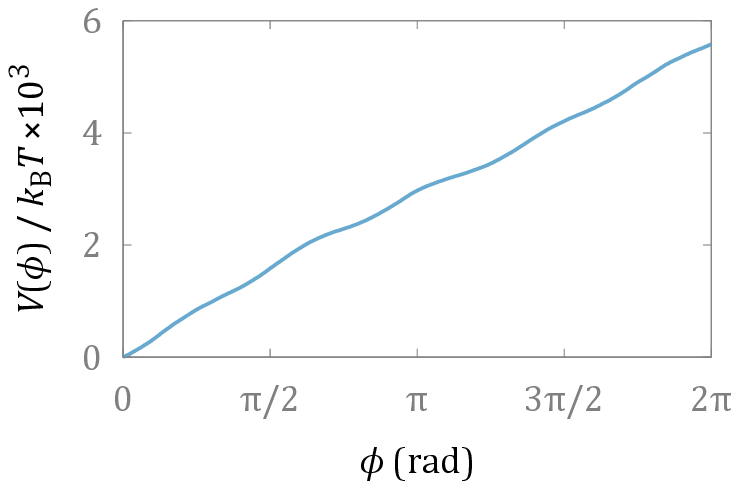}
\end{tabular}
\caption{Conservative (left hand side) and total potentials (right hand side).}\label{fig:potential}
\end{figure}

\noindent We record video of a rotating micro-propeller for a period of 15~minutes, and extract its orientation in each frame as described above. The steady state probability distribution, $p_{\mathrm{ss}}(\phi)$ is constructed as a histogram by binning the available data. We confirm that $p_{\mathrm{ss}}(\phi)$ is a true steady state by comparing with separate orientation distributions derived from the first and second halves of the total Brownian trail. These two distributions are the same, with the exception of a jagged, high frequency component. These sharp features are of very low amplitude in comparison to the overall shape of $p_{\mathrm{ss}}(\phi)$ and derive from multiple sources including the finiteness of the Brownian trail, tracking errors, centre of mass motion and tilting. To prevent unrealistic contributions to the medium entropy introduced via the gradient of $p_{\mathrm{ss}}(\phi)$, we smooth the distribution function. To evaluate $\omega_{\mathrm{ss}}(\phi)$ we form a histogram of the coarse-grained velocity, i.e.\ we take the numerical derivative of the orientation over a single time step and bin the result according to the average orientation over the time step. Figure (\ref{fig:steady}) shows $p_{\mathrm{ss}}(\phi)$, $\omega_{\mathrm{ss}}(\phi)$ and their product, $j_{\mathrm{ss}}$ (see Eq.~(\ref{eq:fp0})). Smoothed data is shown in thin, coloured lines with the raw data plotted on thicker gray lines. The small variation of $j_{\mathrm{ss}}$ with orientation provides a measure of the errors in the experiment. Next, we calculate $V(\phi)$ using the integral in equation~(\ref{eq:V0}) and $\trq=\xi_{\mathrm{r}}\dot{\phi}$ (with $\xi_{\mathrm{r}}=$2.8~aN{\textperiodcentered}m{\textperiodcentered}s/rad). We then perform the arbitrary separation of $V(\phi)$ into a conservative potential, $V^0(\phi)$, and our choice of $T^0$ (the average value of $\trq(\phi)$: $T^0$=-3.59~aN{\textperiodcentered}m), see Fig.~(\ref{fig:potential}).

\noindent Finally, we investigate total change in entropy over different time scales. We choose an interval, $t$, and split the time series of micro-propeller orientations from the experiment into $T/t$ sections, each spanning time $t$. For each of these sections, the total entropy change is calculated from the sum of equations~(\ref{eq:smed}) and (\ref{eq:ssys}), using the end points of the trajectory and the previously calculated values for $p_{\mathrm{ss}}(\phi)$, $T^0$, and $V^0(\phi)$. We repeat this analysis for different intervals, $t$. 

\begin{figure}[h!]
\begin{center}
\includegraphics[width=10cm]{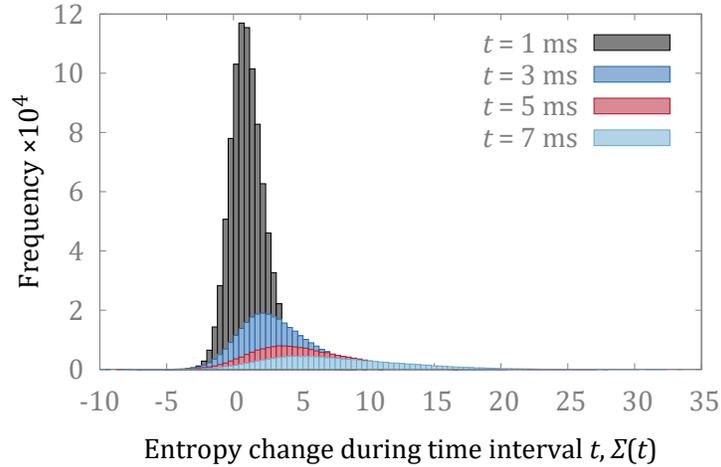}
\end{center}
\caption{Evolution of the distribution of entropy change for a series of increasing time periods.}\label{fig:dst}
\end{figure}

\begin{figure}[h!]
\centering
\begin{tabular}{cc}
\includegraphics[width=8cm]{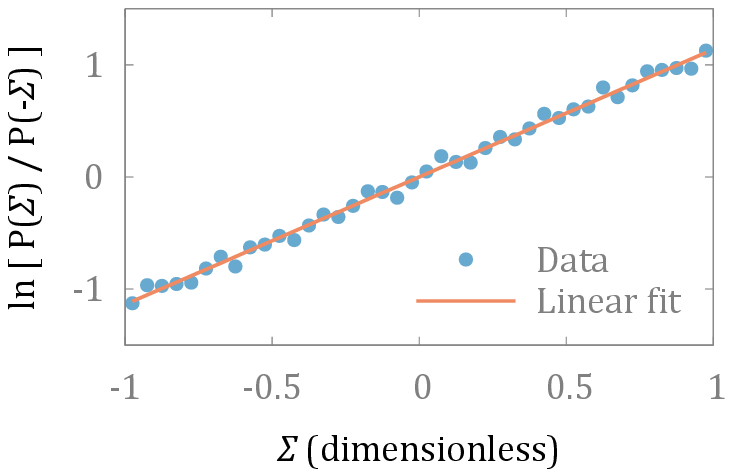}
\includegraphics[width=8cm]{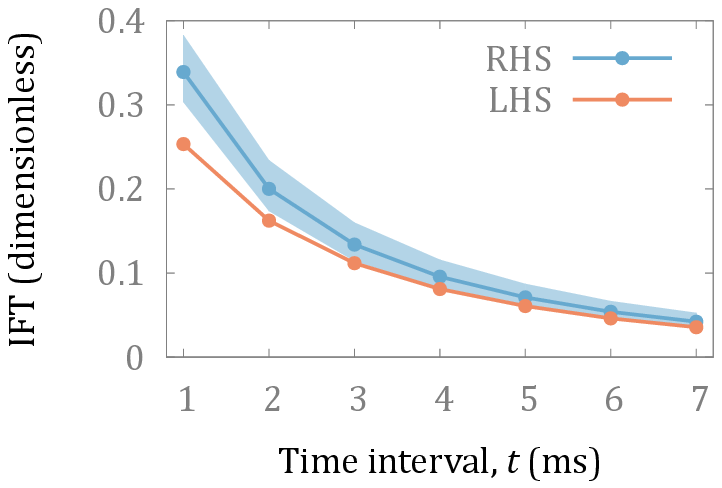}
\end{tabular}
\caption{Experimental tests of the FT. Left hand side: detailed FT plotted as the natural logarithm of Eq. (\ref{eq:dft0}), with best fit straight line with gradient $\approx 1.14$. Right hand side: left and right hand sides of the integrated FT, Eq. (\ref{eq:ift0}).}\label{fig:rft}
\end{figure}

Figure (\ref{fig:dst}) shows histograms of entropy change, $\Sigma_t$, over a sequence of time intervals, $t$. As suggested above, for the shorter time interval, $t=1$~ms, the distribution is strongly peaked close to zero, with a substantial fraction of trajectories showing negative changes in entropy. As the time interval lengthens, the distribution spreads out, and the peak shifts to higher positive values. At each stage the fraction of trajectories showing negative entropy change decreases. This process is quantified by the transient and integrated rotational FTs, which are tested in Fig. (\ref{fig:rft}). On the left hand side, Fig. (\ref{fig:rft}a) we plot the logarithm of Eq.~(\ref{eq:dft0}). In accordance with the FT, the graph is a linear one passing through the origin. The gradient of the best fit line is 1.14, compared with 1, as it would be for perfect agreement. The integrated form is tested by plotting the right and left hand sides of Eq. (\ref{eq:ift0}). The shaded region shows the effect of varying the value of $\xi_{\mathrm{r}}$ by $\pm 15\%$ when evaluating $\trq(\phi)$. Since the rotational friction scales with the cube of the length, this variation corresponds to a change in linear dimension of $\approx \pm 5\%$. The strength of the agreement increases with the time interval, becoming very close for $t \gtrapprox 3$ ms.

\section{Discussion}
We have defined a version of the FT suitable for describing fluctuations of a rudimentary, light driven micro-machine. 

\noindent The scenario we have considered is somewhat contrived: we had to go to considerable lengths to ensure that we could reliably resolve the behaviour of interest. In particular we had to design a propeller that would rotate slowly enough for the transient, entropy consuming trajectories to be accurately measured. The design also minimized motion of the centre of mass, and angular fluctuations of the rotation axis, allowing us to eliminate five of the six physical degrees of freedom and focus on the remaining axial rotations. For the detailed FT, Eq.~(\ref{eq:dft0}) and Fig.~(\ref{fig:rft}a), the experimental data, which are plotted for the logarithm of Eq.~(\ref{eq:dft0}) are represented by a straight line, indicating the exponential dependence expected from the FT, however the gradient is higher than expected. In the case of the integrated FT, Eq. (\ref{eq:ift0}) and Fig. (\ref{fig:rft}b), our measurements tend to under-estimate the entropy changes over very small time intervals, although the agreement becomes more accurate for intervals of longer duration. There are a number of sources of error including image blur and fluctuations in the rotation axis of the micro-propeller. Both of these factors are more significant for short time intervals. In the first case, each video image does not represent an instantaneous moment, but contains contributions from previous times. This has the effect of low pass filtering the measured rotations relative to actual rotations. Tilting of the rotation axis adds a small error to the measured displacement of each tracking sphere, when projected onto a fixed horizontal plane.  In addition the effects of data smoothing (Fig. (\ref{fig:steady})) will be more significant for shorter intervals.

\noindent Nevertheless, the essential features are qualitatively reproduced. For short time intervals, entropy consuming trajectories occur with substantially higher probability, and this probability decreases exponentially as the duration of the time interval increases.

\noindent Although it was not straight forward to observe and quantify this behaviour, we note that fluctuations influence the efficiency and precision of micro-machines whether they are measured or not. The concepts described in this article, and in the many articles devoted to both the fundamental theory, and to applications in biology, are directly relevant to the growing field of artificial micro-machines \cite{Knopf2018Light,Xu2018Micro}. This will become increasingly more true as we try to design increasingly small machines. Indeed, the analysis of the micro-propeller is somewhat simplified by the fact that the underlying potential (Eq. (\ref{eq:V0}), Fig. (\ref{fig:potential})) is many times greater than $k_{\mathrm{B}}T$. For this reason, the effect of diffusion in Eqns. (\ref{eq:jss},\ref{eq:smed}) is relatively weak in comparison to the drift terms produced by the systematic torques. Understanding the role of fluctuations in more complex micro-machines, will present further challenges, especially for machines involving multiple degrees of freedom. Under these circumstances, the steady state distributions are themselves multi-dimensional, complicating the interpretation of force, locally. For instance, if we wish to design a micro-machine that applies a particular force in a particular configuration, we might also need to understand the morphology of $p_{\mathrm{ss}}$ in multiple dimensions. Providing theoretical approximations for the effect of fluctuations on more complex micro-scale devices with many degrees of freedom, including synchronizing systems \cite{Box2015Transitional,Kotar2013Optimal}, presents an interesting challenge and one that could be technologically useful. One possibility would be to try to eliminate spurious degrees of freedom, by considering the cycle of the machine as a one dimensional curve embedded in a higher dimensional phase space. In fact, this is precisely what has been done for the rotor considered above. Ultimately, understanding the interplay between deterministic and fluctuating forces is essential for the design and optimization of complex, multi-dimensional, mesoscopic machines. 




\bibliography{rft_mpa}

\end{document}